\documentclass[aps,pre,floatfix]{revtex4} 
\usepackage{epsf} 
\usepackage{subfigure} 
\usepackage{amsmath} 
\usepackage{amssymb} 
\usepackage{graphicx} 
\usepackage{epstopdf} 
\begin{document}

\newcommand{\be}{\begin{equation}}
\newcommand{\ee}{\end{equation}}
\newcommand{\bea}{\begin{eqnarray}}
\newcommand{\eea}{\end{eqnarray}}

\title{\bf Discrete- versus continuous-state descriptions of the F$_{1}$-ATPase molecular motor} 

\author{E. Gerritsma and P. Gaspard} 
\affiliation{Center for Nonlinear Phenomena and Complex Systems,\\ 
Universit\'e Libre de Bruxelles, Code Postal 231, Campus Plaine, 
B-1050 Brussels, Belgium} 

\begin{abstract}
A discrete-state model of the F$_{1}$-ATPase molecular motor is developed
which describes not only the dependences of the rotation and ATP consumption rates
on the chemical concentrations of ATP, ADP, and inorganic phosphate, but also
on mechanical control parameters such as the friction coefficient and the external torque.
The dependence on these mechanical parameters is given to the discrete-state model
by fitting its transition rates to the continuous-angle model of P. Gaspard and E. Gerritsma [J. Theor. Biol. {\bf 247} (2007) 672-686].  This discrete-state model describes the behavior of the F$_1$ motor
in the regime of tight coupling between mechanical motion and chemical reaction.  In this way,
kinetic and thermodynamic properties of the F$_1$ motor are obtained such as the Michaelis-Menten dependence of the rotation and ATP consumption rates on ATP concentration and its extension in the presence of ADP and P$_{\rm i}$, their dependences on friction and external torque, as well as the chemical and mechanical thermodynamic efficiencies.

\vskip 0.5 cm

{\it Keywords:} molecular motor, F$_1$-ATPase, mechanochemical coupling, stochastic process, nonequilibrium thermodynamics.

\end{abstract}

%\noindent Journal of Theoretical Biology {\bf xxx} (200x) xxx-xxx

%\vskip 0.5 cm

\maketitle

\section{Introduction}

F$_{\rm o}$F$_{1}$-ATPase is a ubiquitous protein producing adenosine triphosphate (ATP) in mitochondria \cite{Alberts,oster}. {\it In vivo}, the F$_{\rm o}$ part of this protein is embedded in the inner membrane of mitochondria and is rotating as a turbine when a proton current flows through it, across the membrane.  This turbine drives the rotation of a shaft inside the hydrophylic F$_1$ part.
This latter is composed of three $\alpha$- and three $\beta$-subunits spatialy alternated as a hexamer $(\alpha\beta)_3$ and forming a barrel for the rotation of the shaft made of a $\gamma$-subunit \cite{Abrahams,jewalker}.
Upon rotation, the central $\gamma$-shaft induces conformational changes in the hexamer, leading to the synthesis of ATP in catalytic sites located in each $\beta$-subunit.  In their experimental work, Kinosita and coworkers have succeeded to build a nanomotor by separating the F$_1$ part and attaching an actin filament or a colloidal bead to its $\gamma$-shaft. {\it In vitro}, ATP hydrolysis drives the rotation of this nanomotor, transforming chemical free energy from ATP into the mechanical motion of the $\gamma$-shaft \cite{N386,cell2007}. This motion proceeds in steps of $120^{\circ}$, revealing the three-fold symmetry of F$_1$-ATPase \cite{Abrahams,jewalker,N386}.
Furthermore, the experiments have shown that each of these steps is subdivided into two substeps \cite{N410,N427}.
The first substep of about $90^{\circ}$ is attributed to ATP binding to an empty catalytic $\beta$-subunit of F$_{1}$, followed by conformational changes of this subunit and of the whole hexamer as a consequence of its binding to the other subunits.  This deformation of the protein generates a torque on the $\gamma$shaft, hence its rotation by about $90^{\circ}$. The secondary substep of about $30^{\circ}$ follows the hydrolysis of ATP and results into the release of the products adenosine diphosphate (ADP) and inorganic phosphate (P$_{\rm i}$), completing the 120$^{\circ}$ step \cite{cell2007,pnas2003}.

In our previous work \cite{pgeg}, we carried out a theoretical study of the stochastic chemomechanics of the F$_1$-ATPase molecular motor on the basis of the experimental observations reported in Ref. \cite{N410}. The stochasticity of the motion is the consequence of the nanometric size of the F$_1$ motor making it sensitive to the thermal and molecular fluctuations due to the atomic structure of the protein and surrounding medium \cite{NP77,G04}.  In our paper \cite{pgeg}, the rotation angle was taken as a continuous random variable and the stochastic process ruled by six coupled Fokker-Planck equations for the biased diffusion of the angle and the random jumps between the six chemical states considered in our model. This continuous-angle model allows us to reproduce accurately the experimental observations of Ref. \cite{N410} and, in particular, the random trajectories of the $\gamma$-shaft  with the steps and also the substeps \cite{pgeg}.  Taking the angle as a continuous variable provides a realistic description of the random motion of the motor.  

However, coarser descriptions are often considered in which the angle (or the position in the case of linear motors) performs discrete jumps instead of varying continuously. The discretization is considered not only because it is always required in order to simulate the random process of diffusion for the angle \cite{XWO05}, but also because the angle is observed to jump by finite amounts
corresponding to the steps and substeps observed in the experiments \cite{N410,N427}.
This observation suggests an alternative description with discrete states associated 
with the steps and substeps.  As long as the steps and substeps can be identified with the chemical states of the rotary motor, this fully discrete description can be set up on the basis of chemical kinetics. 
This description could in principle be deduced by coarse graining the continuous-angle description,
but is most often taken as a model with parameters to be fitted to experimental observations. 
In such coarse-grained descriptions in terms of discrete states, the stochastic process is no longer a process with both diffusion and jumps, but becomes a purely jump process ruled by a so-called master equation \cite{NP77,G04}.  Such discrete-state descriptions provide simpler models of molecular motors, relying on similar but different assumptions compared to the finer descriptions using continuous random variables for angle or position beside the discrete chemical states.  
In this context, the question arises of understanding the relationship between the two levels of description of the molecular motor. Since both discrete- and continuous-state models of molecular motors have been proposed in the literature, it is becoming important to develop methods to relate both levels of description, which is the purpose of the present paper.

The comparison between the two levels of description is interesting not only for the methodology, but also because it can reveal important properties of the molecular motor such as the nature of the coupling between their chemistry and mechanics, i.e., the question whether the mechanochemical coupling is tight or loose. As discussed in the following, this property can be expressed in terms of the nonequilibrium thermodynamics of the molecular motor.

The paper is organized as follows.
In section \ref{Discrete}, we present in detail the discrete-state description of the F$_1$ molecular motor.
In section \ref{Continuous}, this description is compared with our previous model with a continuous angle variable.  In section \ref{Properties}, we discuss the properties of the F$_{1}$ motor in the light of the comparison between the discrete- and continuous-state descriptions and, in particular, the question of the mechanochemical coupling and its thermodynamic efficiencies.
The conclusions are drawn in section \ref{Conclusions}.

\section{Discrete-state description}
\label{Discrete}

\subsection{Chemistry of the F$_1$ motor}

In the discrete-state description, the discrete values of the angle of the $\gamma$-shaft correspond to the chemical states so that the mechanical motion of the motor is directly controlled by its chemistry.

The motor is powered by the hydrolysis of ATP into ADP and P$_{\rm i}$:
\begin{equation}
{\rm ATP} \rightleftharpoons {\rm ADP}+{\rm P}_{\rm i}
\label{atp}
\end{equation}
This reaction is driven by the difference of chemical potential $\Delta \mu$ between the three species (ATP, ADP, and P$_{\rm i}$):
\begin{eqnarray}
\Delta \mu = \mu_{\rm ATP}-\mu_{\rm ADP}-\mu_{\rm P_{i}}
\label{deltamuATP}
\end{eqnarray}
where the chemical potential $\mu_{\rm X}$ is equal to the corresponding Gibbs free energy per molecule defined as
\begin{eqnarray}
\mu_{\rm X}=\mu_{\rm X}^{0}+k_{\rm B}T\ln \frac{[{\rm X}]}{c^{0}}
\label{defPotchim}
\end{eqnarray}
with X = ATP, ADP, or P$_{\rm i}$, the absolute temperature $T$, Boltzmann's constant $k_{\rm B}=1.38 \, 10^{-23}$ J/K, and the reference concentration $c^0=1$ mole per liter at which the chemical potential of species X takes its standard value $\mu_{\rm X}^{0}$.

The standard Gibbs free energy of ATP hydrolysis takes the value $\Delta G^{0}=-\Delta \mu^0=-30.5$ kJ/mol $= -7.3$ kcal/mol $= -50$~pN~nm at the temperature of $23^{\circ}$C, the external pressure of $1$~atm, and pH~$7$ \cite{KAI04}. We notice that ATP hydrolysis provides a significant amount of free energy of $-\Delta G^{0}=12.2 \, k_{\rm B}T$ above the thermal energy $k_{\rm B}T= 4.1$~pN~nm.
Since the chemical potential difference (\ref{deltamuATP}) vanishes at equilibrium,
the equilibrium concentrations of ATP, ADP, and P$_{\rm i}$ satisfy
\begin{equation}
\frac{[{\rm ATP}]}{[{\rm ADP}][{\rm P_{i}}]}\Biggl\vert_{\rm eq} = \exp \frac{\Delta G^{0}}{k_{\rm B}T}
\simeq 4.9\, 10^{-6} \, {\rm M}^{-1}
\label{eq}
\end{equation}
showing that ATP tends to hydrolyze into its products. In Eq. (\ref{eq}) and the following, we no longer write the reference concentration $c^0$, assuming that the concentrations are counted in mole per liter (M).

The kinetic scheme of our model is based on the phenomenological observations of 120$^{\circ}$ rotation of the $\gamma$-shaft per consumed ATP molecule.
In accordance with Ref. \cite{N410}, the first substep, the 90$^{\circ}$ rotation of the $\gamma$-shaft,  is induced by the binding of ATP to an empty catalytic site.
The second substep, the 30$^{\circ}$ rotation of the $\gamma$-shaft, is induced by the release of ADP and P$_{i}$. The process can be summarized by the following chemical scheme

\begin{eqnarray}
{\rm ATP}+\underbrace{[\emptyset,\gamma(\theta)] }_{\rm state\, 1} \;
\mathop{\rightleftharpoons}^{W_{+1}}_{W_{-1}} \;
\underbrace{[{\rm ATP}^{\ddagger},\gamma(\theta+90^{\circ})]}_{\rm state \, 2}\; \mathop{\rightleftharpoons}^{W_{+2}}_{W_{-2}}\; \underbrace{[\emptyset,\gamma(\theta+120^{\circ})]}_{\rm state\, 1}+{\rm ADP}+{\rm P_{i}}
\label{schemaCin}
\end{eqnarray}
From left to right: In state 1, ATP can bind to an empty ($\emptyset$) $\beta$-catalytic site of F$_{1}$ with the $\gamma$-shaft at angular position $\theta$.
State 1 is thus defined by $[\emptyset,\gamma(\theta)]$.
Binding of ATP induces the $90^{\circ}$ rotation of the $\gamma$-shaft, which we represent by $\gamma(\theta+90^{\circ})$ and fills this catalytic site.
ATP$^{\ddagger}$ stands for any transition state of ATP between the initial triphosphate molecule to the products of hydrolysis ADP and P$_{\rm i}$ before the evacuation of the $\beta$-catalytic site.
State 2 is thus represented by $[{\rm ATP}^{\ddagger},\gamma(\theta+90^{\circ})]$.
If F$_{1}$ proceeds to hydrolysis, the products ADP and P$_{\rm i}$ are released together, which induces the secondary 30$^{\circ}$ rotation and empties a $\beta$-subunit.

Notice that Eq. (\ref{schemaCin}) does not necessarily represent the same $\beta$-subunit during different \textit{real} catalytic states.
Following our reference data \cite{N410}, ADP and P$_{\rm i}$ are released together, but from a different $\beta$-subunit then the ATP binding one.
In this case, Eq. (\ref{schemaCin}) represents different $\beta$-subunits; one that binds ATP and one that releases ADP and P$_{\rm i}$ from another ATP molecule previously bound to an other $\beta$-catalytic subunit.  Furthermore, if the concentration [ATP] is below the nanomolar, F$_{1}$ can perform uni-site catalysis \cite{oneMeca}. In this other case, Eq. (\ref{schemaCin}) would represent a single $\beta$-subunit in its two main catalytic states.   The relevance of Eq. (\ref{schemaCin}) to both cases is a consequence of our approach which consists in looking only at the contributions of ATP hydrolysis to the rotation of $\gamma$. Very recent experimental observations have shown that ADP and P$_{\rm i}$ may be released from different $\beta$-subunits \cite{cell2007}.
If both product molecules are released at different times, a more complete model could be considered by adding a third state to the kinetic scheme.  Here, the scheme is simplified by lumping together the two states subsequent to hydrolysis in a regime where the time delay between the releases of ADP and P$_{\rm i}$ is short enough, as also considered elsewhere \cite{pgeg}.

Since F$_1$ is a hexamer composed of three $\beta$-subunits, the reactions (\ref{schemaCin}) appear three times for the angles $\theta+ 120^{\circ} n$ with $n=0,1,2$, so that the motor has a total of six chemical states.  However, by the three-fold symmetry of the F$_1$ motor, these six states can be regroup three by three since the process repeats itself similarly every $120^{\circ}$ in each $\beta$-subunit.  We can therefore identify the three states of type 1 together in state 1 and the same for state 2, reducing the motor dynamics to a process with only two discrete states.

According to the mass-action law of chemical kinetics, the reaction rates $W_{\rho}$ in Eq. (\ref{schemaCin}) depend on the molecular concentrations in the solution surrounding the motor as follows
\begin{eqnarray}
\label{tautr+1}
W_{+1}&=&k_{+1}[{\rm ATP}]\\
\label{tautr-1}
W_{-1}&=&k_{-1}\\
\label{tautr+2}
W_{+2}&=&k_{+2}\\
\label{tautr-2}
W_{-2}&=&k_{-2}[{\rm ADP}][{\rm P_{i}}]
\end{eqnarray}
where the quantities $k_{\rho}$ $(\rho=\pm1, \pm2)$ are the constants of the forward and backward reactions of binding and unbinding of ATP or ADP with P$_{\rm i}$ and, [ATP], [ADP], [P$_{\rm i}$] represent the concentrations of each species. $k_{+1}$ is the constant of ATP binding often denoted $k_{\rm on}$ and $k_{-1}$ the ATP unbinding constant $k_{\rm off}$, while $k_{+2}$ is the constant of ATP catalysis denoted $k_{\rm cat}$. The two-state model includes the possibility of ADP and P$_{i}$ binding and thus ATP synthesis, with the corresponding kinetic constant $k_{-2}$.
Since the rates $W_{\rho}$ are measured in s$^{-1}$, the constant $k_{+1}$ has the units of 
M$^{-1}$s$^{-1}$, $k_{-1}$ and $k_{+2}$ the units of s$^{-1}$, and $k_{-2}$ the units of M$^{-2}$s$^{-1}$.

An important aspect of the experiments with beads or actin filaments attached to the $\gamma$-shaft is that the behavior of the molecular motor also depends on the friction $\zeta$ of the attached objects moving in the viscous medium surrounding F$_1$, as well as on the external torque $\tau$ which is applied in some experiments \cite{N427,N433}.  As a consequence, the reaction constants (\ref{tautr+1})-(\ref{tautr-2}) depend on both the friction $\zeta$ and the external torque $\tau$.
Indeed, a discrete-state description in which the motion is directly coupled to the chemical reactions leaves no place to the modeling of mechanical aspects such as friction and external torque.  On the other, a continuous-angle description based on a stochastic Newtonian equation of Langevin type would provide a way to include the mechanical forces of friction and external torque by using the laws of Newtonian mechanics.  This was achieved in our previous paper \cite{pgeg}, allowing us to study the dependence on friction and external torque.  In order to take into account these mechanical aspects in the present discrete-state model, we have to incorporate the effect of friction $\zeta$ and external torque $\tau$ into the reaction constants $k_{\rho}(\zeta,\tau)$ ($\rho = \pm1, \pm2$) and to find these dependences by comparison with the continuous-angle model \cite{pgeg} which includes the mechanical aspects. This will be achieved in section \ref{Continuous} where the discrete-state model will be completed in this way. We notice that a discrete-state model with six chemical states has been studied in Ref. \cite{AG06} in the particular case where the external torque is vanishing and for a fixed value of the friction coefficient.  Here, our purpose is to investigate situations where the molecular motor is submitted to an external torque to understand how the coupling between chemistry and mechanics can be formulated in terms of the reaction rates (\ref{tautr+1})-(\ref{tautr-2}) of the discrete-state model.

\subsection{Thermodynamics of the F$_1$ motor}

In order for the description to be consistent with thermodynamics, let us summarize here the conditions that the chemical and mechanical properties of the motor must satisfy.

The internal energy of the motor changes with the rotation angle $\theta$ and the numbers $N_{\rm X}$ of molecules (X=ATP, ADP, P$_{\rm i}$) entering the catalytic sites according to the Gibbs relation
\be
dE= T \, dS + \tau \, d\theta+ \sum_{\rm X} \mu_{\rm X} \, dN_{\rm X}
\label{Gibbs}
\ee 
where $\tau$ denotes the external torque, $\mu_{\rm X}$ the chemical potential (\ref{defPotchim}), $T$ the temperature, and $S$ the thermodynamic entropy.  The changes in molecular numbers due to the overall reaction (\ref{atp}) satisfy
\be
dN_{\rm ATP} = - dN_{\rm ADP} = -dN_{\rm P_{i}}
\ee
where $dN_{\rm X}$ is the number of molecules of species X entering the motor.  Consequently, the Gibbs relation (\ref{Gibbs}) becomes
\be
dE= T \, dS + \tau \, d\theta+\Delta\mu \, dN_{\rm ATP}
\label{Gibbs2}
\ee 
in terms of the chemical potential (\ref{deltamuATP}) of the overall reaction.
In an open system such as a motor, energy and entropy vary {\it a priori} because of exchanges with the surrounding medium or possibly due to internal variations:
\bea
dE &=& d_{\rm e}E + d_{\rm i}E \\
dS &=& d_{\rm e}S + d_{\rm i}S
\eea
The laws of thermodynamics only rule the internal variations:
\bea
\mbox{first law:}\quad && d_{\rm i}E=0 \label{1st_law}\\
\mbox{second law:}\quad && d_{\rm i}S \geq 0 \label{2nd_law}
\eea
Since energy and entropy are state variables, they recover their starting value after completing a cyclic process in a motor whereupon their integration over a cycle vanishes, $\oint dE=0$ and $\oint dS =0$. For the isothermal process of a nanomotor in the heat bath given by the surrounding medium, the entropy irreversibility produced over a motor cycle is thus given by
\be
\oint d_{\rm i} S = - \oint d_{\rm e} S = \frac{1}{T} \oint \left( \tau \, d\theta + \Delta\mu \, dN_{\rm ATP}\right) \geq 0
\ee
which is deduced by using Eqs. (\ref{Gibbs2})-(\ref{2nd_law}) for a cyclic process.
Introducing the mean angular velocity of the motor in revolution per second
\be
V \equiv \frac{1}{2\pi} \left\langle \frac{d\theta}{dt}\right\rangle
\ee
and the mean rate of ATP consumption 
\be
R \equiv \left\langle \frac{dN_{\rm ATP}}{dt}\right\rangle
\ee
the entropy production is given by
\be
\frac{d_{\rm i} S}{dt} = \frac{2\pi\tau}{T} \, V + \frac{\Delta\mu}{T} \, R \geq 0
\label{diSdt}
\ee
in terms of the so-called thermodynamics forces or affinities, $2\pi\tau/T$ and $\Delta\mu/T$, and the corresponding fluxes or currents, $V$ and $R$ \cite{P67,KP98,Schnak}.  All these quantities vanish at the thermodynamic equilibrium.  Mechanical or chemical energies are provided from the exterior if either the torque $\tau$ or the reaction free energy $\Delta\mu$ are non vanishing.

In general, molecular motors can be driven out of equilibrium by either the external torque $\tau$ or the molecular concentrations of ATP, ADP and P$_{\rm i}$ entering $\Delta\mu$, possibly in combination.  In this regard, the thermodynamic forces $2\pi\tau/T$ and $\Delta\mu/T$ are independent control parameters.  However, in a regime where the chemistry and the mechanics of the molecular motor are tightly coupled, three ATP molecules are consumed per revolution which is expressed by the conditions that the ATP consumption rate is three times the velocity, i.e.,
\be
\mbox{tight coupling:} \qquad V = \frac{1}{3} \, R
\label{V_tc}
\ee
In this case, the entropy production (\ref{diSdt}) becomes
\be
\mbox{tight coupling:} \qquad \frac{1}{k_{\rm B}} \frac{d_{\rm i} S}{dt} = A R \geq 0
\label{diSdt_tc}
\ee
with the chemomechanical affinity
\begin{equation}
A \equiv \underbrace{\frac{2\pi}{3}\frac{\tau}{k_{\rm B}T}}_{\rm mechanics} + \underbrace{\frac{\Delta \mu}{k_{\rm B}T}}_{\rm chemistry}
\label{aff}
\end{equation}
which is here written in dimensionless form.  In the regime of tight coupling, the mechanical and chemical thermodynamic forces are thus no longer independent control parameters but are replaced by the unique chemomechanical affinity (\ref{aff}).  This affinity vanishes at equilibrium.  By using Eq. (\ref{defPotchim}), we obtain the equilibrium condition for the concentrations:
\be
\frac{[{\rm ATP}]}{[{\rm ADP}][{\rm P_{i}}]}\Biggl\vert_{\rm eq} 
= \exp \frac{1}{k_{\rm B}T}\left(\Delta G^{0}-\frac{2\pi}{3}\tau \right) 
\simeq 4.9\, 10^{-6} {\rm M}^{-1} \exp\left( -\frac{2\pi}{3}\frac{\tau}{k_{\rm B}T}\right)
\label{eq+torque}
\ee
in the presence of the external torque $\tau$. This result shows that the chemical equilibrium is displaced in the presence of an external torque.  {\it In vivo}, such an external torque comes from the F$_{\rm o}$ part of ATPase and the transmembrane pH difference in mitochondria.  This displacement of equilibrium finds its origin in the mechanochemical coupling achieved in the F$_1$ rotary motor but happens only if an external torque is enforced on the $\gamma$-shaft.

In the discrete-state description, the discrete values of the angle of the $\gamma$-shaft correspond to the chemical states so that the mechanical motion is tightly coupled to the chemistry of F$_1$, a feature which is characteristic of such discrete models.  Therefore, the relations (\ref{V_tc}), (\ref{diSdt_tc}), and (\ref{aff}) apply to such tight-coupling models. 

\subsection{Master equation description}
\label{master}

The chemistry of F$_{1}$ is a stochastic process because the arrival of each substrate molecule (ATP or ADP with P$_{\rm i}$) is a random event in time.  Accordingly, the time evolution of the molecular motor is described in terms of the probabilities $P_{\sigma}(t)$ to find it in one or the other of the two states $\sigma=1,2$ of the kinetic scheme (\ref{schemaCin}).  These probabilities evolve in time because of the random transitions between both states at each reaction.  This time evolution is thus ruled by the master equation \cite{NP77,G04}:
\begin{equation}
\frac{d P_{\sigma}(t)}{dt} = \sum_{\rho ,\sigma'} \left[ 
W_{\rho}(\sigma'\vert \sigma) P_{\sigma'}(t)
- W_{-\rho}(\sigma \vert \sigma')P_{\sigma}(t) \right]
\label{meq}
\end{equation}
with a sum over the two reactions $\rho=1,2$ and the two states $\sigma'=1,2$ before the transition $\sigma' {\overset{\rho}\longrightarrow} \sigma$ or after the reverse transition $\sigma {\overset{-\rho}\longrightarrow} \sigma'$. The quantity $W_{\rho}(\sigma'\vert \sigma)$ is the transition rate per unit time from the state $\sigma'$ to the state $\sigma$ due to the reaction $\rho$, which can be identified with the rate of the corresponding reaction between  the two chemical states $\sigma$ and $\sigma'$. The master equation conserves the total probability $\sum_{\sigma}P_{\sigma}(t)=1$ for all times $t$.
Using the specific values of the four reaction rates of Eq. (\ref{schemaCin}) resumed in Eqs. (\ref{tautr+1})-(\ref{tautr-2}), we write the time evolution of the states 1 and 2 with respectively empty and occupied catalytic site as
\begin{eqnarray}
\frac{dP_{1}}{dt}&=&(W_{-1}+W_{+2})P_{2}-(W_{+1}+W_{-2})P_{1}\\
\frac{dP_{2}}{dt}&=&(W_{+1}+W_{-2})P_{1}-(W_{-1}+W_{+2})P_{2}
\label{probaevo}
\end{eqnarray}
with the normalisation condition $P_1+P_2=1$ always satisfied.  

The mean consumption rates of the different species are given by
\bea
\frac{d}{dt}\langle N_{\rm ATP} \rangle &=& W_{+1} P_1 - W_{-1} P_2 \\
\frac{d}{dt}\langle N_{\rm ADP} \rangle = \frac{d}{dt}\langle N_{\rm P_i} \rangle &=& W_{-2} P_1 -W_{+2} P_2
\eea
Since the $\gamma$-shaft rotates by a substep of $90^{\circ}$ during the first reaction $\rho=+1$ and by a substep of $30^{\circ}$ during the second reaction $\rho=+2$, the mean angle of the $\gamma$-shaft 
evolves in time according to
\be
\frac{d}{dt}\langle\theta\rangle= \frac{\pi}{2} \left( W_{+1} P_1 - W_{-1} P_2 \right) + \frac{\pi}{6} 
\left( W_{+2} P_2 - W_{-2} P_1 \right)
\label{ang}
\ee
in radian per second.  

{\it Stationary state.}  The master equation admits a time-independent stationary solution such that $(d/dt)P_{\sigma}^{\rm (st)}=0$.
The stationary solution for each two states is given by
\begin{eqnarray}
P_{1}^{\rm (st)}&=&\frac{k_{-1}+k_{+2}}{k_{+1}[{\rm ATP}]+k_{-1}
+k_{+2}+k_{-2}[{\rm ADP}][{\rm P_i}]} \label{etatstat1}\\
P_{2}^{\rm (st)}&=&\frac{k_{+1}[{\rm ATP}]+k_{-2}[{\rm ADP}][{\rm P_i}]}{k_{+1}[{\rm ATP}]+k_{-1}
+k_{+2}+k_{-2}[{\rm ADP}][{\rm P_i}]}
\label{etatstat2}
\end{eqnarray}
where we used the normalization condition $P_{1}^{\rm (st)}+P_{2}^{\rm (st)}=1$.
We notice that this state is stationary in the statistical sense because an individual motor continues to fluctuate in time.  The probabilities (\ref{etatstat1})-(\ref{etatstat2}) define the statistical distribution of its fluctuations between the states 1 and 2 after sampling over a long enough lapse of time and over many random realizations.

In the stationary state, the mean angular velocity of the $\gamma$-shaft is thus given by
\be
V = \frac{1}{2\pi} \frac{d}{dt}\langle \theta\rangle_{\rm st}= \frac{1}{3}\frac{k_{+1}k_{+2}[{\rm ATP}]-k_{-1}k_{-2}[{\rm ADP}][{\rm P_i}]}{k_{+1}[{\rm ATP}]+k_{-1}+k_{+2}+k_{-2}[{\rm ADP}][{\rm P_i}]}
\label{v}
\ee
and the mean rates of molecular consumption by
\bea
R = \frac{d}{dt}\langle N_{\rm ATP} \rangle_{\rm st} = -
\frac{d}{dt}\langle N_{\rm ADP} \rangle_{\rm st} = -\frac{d}{dt}\langle N_{\rm P_i} \rangle_{\rm st}= 3V
\label{R}
\eea
The condition (\ref{V_tc}) of tight chemomechanical coupling is thus satisfied.

{\it Equilibrium state.} At thermodynamical equilibrium, 
all the nonequilibrium constraints vanish.  In this case, the stationary solution represents the equilibrium probability $P_{\sigma}^{\rm (eq)}$ and obeys the detailed balance conditions \cite{pierr}
\be
W_{\rho}(\sigma'\vert \sigma) P_{\sigma'}^{\rm (eq)}=W_{-\rho}(\sigma \vert \sigma')P_{\sigma}^{\rm (eq)}
\label{detbal}
\ee
for all reactions $\rho=1,2$ and all the transitions $\sigma,\sigma'=1,2$. As a consequence, the mean velocity (\ref{v}) and the rate (\ref{R}) vanish which occurs under the condition:
\be
\frac{[{\rm ATP}]}{[{\rm ADP}][{\rm P_i}]}\Biggl\vert_{\rm eq}=\frac{k_{-1}k_{-2}}{k_{+1}k_{+2}} \label{ratioFb}
\ee
Comparing with the condition obtained from the thermodynamics of the motor, we get the equilibrium constant of the reaction:
\be
K_{\rm eq} \equiv \frac{k_{-1}k_{-2}}{k_{+1}k_{+2}}
= \exp \frac{1}{k_{\rm B}T}\left(\Delta G^{0}-\frac{2\pi}{3}\tau \right) 
\simeq 4.9\, 10^{-6} {\rm M}^{-1} \exp\left( -\frac{2\pi}{3}\frac{\tau}{k_{\rm B}T}\right)
\label{K_eq}
\ee
which depends on the external torque $\tau$ acting on the molecular motor.  
We see that considerations of equilibrium thermodynamics give a constraint allowing us to fix one reaction constant if we know the three other constants.
However, equilibrium thermodynamics does not provide enough relations in order to determine the four reaction constants.  We have therefore to use data from the kinetics of the motor, i.e., when the motor is out of equilibrium.  This is the purpose of the following subsection.

\subsection{Determination of the reaction constants}
\label{derivConst}

The motor is in a nonequilibrium stationary state if the chemomechanical affinity (\ref{aff}) is non-vanishing, i.e., if the equilibrium condition (\ref{ratioFb}) is not satisfied.  In this case, the mean velocity (\ref{v}) does not vanish and can be compared with experimental data \cite{N410} or with the continuous-angle model \cite{pgeg} which is complete with respect to the mechanics.
In this way, each one of the reaction constants $k_{\rho}$ can be determined by appropriate limits
at fixed values of $\zeta$ and $\tau$, as explained here below.

First of all, the mean velocity (\ref{v}) can be rewritten in the following form:
\be
V = \frac{ V_{\rm max} \, \left( [{\rm ATP}]-K_{\rm eq}[{\rm ADP}][{\rm P}_{\rm i}]\right)}{[{\rm ATP}] + K_{\rm M}+ K_{\rm P}[{\rm ADP}][{\rm P}_{\rm i}]}
\label{ATP+ADP.kinetics}
\ee
in terms of the constants
\bea
V_{\rm max} &\equiv& \frac{1}{3}\; k_{+2}
\label{v_max} \\
K_{\rm M} &\equiv& \frac{k_{-1}+k_{+2}}{k_{+1}}
\label{K_M}\\
K_{\rm P}  &\equiv& \frac{k_{-2}}{k_{+1}}
\label{K_P}
\eea
together with the equilibrium constant $K_{\rm eq}$ defined by Eq. (\ref{K_eq}).
The knowledge of these four constants is equivalent to knowing the four reaction constants $k_{\pm 1}$ and $k_{\pm 2}$.  Each one of the three constants (\ref{v_max}), (\ref{K_M}), and (\ref{K_P}) can be determined in a specific regime of functioning of the molecular motor.

The constants (\ref{v_max}) and (\ref{K_M}) can in particular be determined in the absence of the products of hydrolysis, $[{\rm ADP}][{\rm P}_{\rm i}]=0$, in which case the velocity follows a typical Michaelis-Menten kinetics:
\be
V = \frac{V_{\rm max}[{\rm ATP}]}{[{\rm ATP}]+K_{\rm M}}
\label{v_MM2}
\ee
Here, we see that Eq. (\ref{K_M}) is the Michaelis-Menten constant defined as the ATP concentration at which the velocity $V$ equals $V_{\rm max}/2$ \cite{HMSPR05}.  This constant
characterizes the crossover between the regime $[{\rm ATP}]\ll K_{\rm M}$ where the rotation is limited by the slow arrival of ATP molecules and the saturation regime $[{\rm ATP}]\gg K_{\rm M}$ 
where the velocity reaches its maximum value fixed by the finite rate of release of ADP and P$_{\rm i}$.  Accordingly, the constant (\ref{v_max}) is the maximum angular velocity of the motor, which is reached in a regime where the ATP concentration is large enough with respect to the Michaelis-Menten constant:
\be
V_{\rm max}  = \lim_{[{\rm ATP}]\to\infty} V
\label{v_max.2}
\ee
the limit meaning that $[{\rm ATP}]\gg K_{\rm M}$.
The Michaelis-Menten constant itself can be determined at low ATP concentrations according to
\be
K_{\rm M}=\lim_{[{\rm ATP}]\to 0} [{\rm ATP}]\left( \frac{V_{\rm max}}{V}-1\right)
\ee
this other limit meaning that $[{\rm ATP}]\ll K_{\rm M}$.

The third constant (\ref{K_P}) characterizes the decrease of the velocity as the concentrations of the products ADP and P$_{\rm i}$ are increased.  It turns out as we shall see in the following that this constant is larger by several orders of magnitude than the equilibrium constant:
\be
K_{\rm P} \gg K_{\rm eq}
\label{P>eq}
\ee
Accordingly, the term involving the equilibrium constant in the numerator of Eq. (\ref{ATP+ADP.kinetics}) can be neglected. Hence, the constant $K_{\rm P}$ can be determined in the regime where
\be
\frac{[{\rm ATP}]+K_{\rm M}}{K_{\rm P}} \ll [{\rm ADP}]_{1,2}[{\rm P}_{\rm i}] \ll \frac{[{\rm ATP}]}{K_{\rm eq}}
\ee
by taking the difference of the inverses of the velocities at two different concentrations of ADP, keeping fixed the other concentrations:
\be
K_{\rm P} = V_{\rm max} \frac{[{\rm ATP}]}{[{\rm P}_{\rm i}]} \frac{\Delta (1/V)}{\Delta[{\rm ADP}]}
\ee
with the notation $\Delta X= X_2-X_1$.

Another consequence of the inequality (\ref{P>eq}) is obtained by using the definition (\ref{K_eq}) of the equilibrium constant:
\be
k_{-1} = \frac{K_{\rm eq}}{K_{\rm P}} \, k_{+2} \ll k_{+2}
\ee
whereupon the Michaelis-Menten constant (\ref{K_M}) is essentially independent of $k_{-1}$ in the present case:
\be
K_{\rm M} = \frac{k_{+2}}{k_{+1}}
\ee
and thus directly determines the constant $k_{+1}$ of ATP binding once the constant $k_{+2}$ of product release is obtained thanks to the maximum velocity (\ref{v_max}).  Subsequently, the constant $k_{-2}$ of the binding of ADP and P$_{\rm i}$ is determined from the constant $K_{\rm P}$ as explained here above.  Finally, the constant $k_{-1}$ of ATP unbinding is determined with the equilibrium constant (\ref{K_eq}).  In summary, we find successively:
\bea
k_{+2} &=& 3 \, V_{\rm max} \label{k+2}\\
k_{+1} &=& \frac{k_{+2}}{K_{\rm M}} \label{k+1}\\
k_{-2} &=& k_{+1} K_{\rm P} \label{k-2}\\
k_{-1} &=& \frac{k_{+1}k_{+2}}{k_{-2}} \, K_{\rm eq} \label{k-1}
\eea

\section{Comparison with the continuous-state description}
\label{Continuous}

In this section, we compare the aforementioned discrete-state description with a continuous-state description we previously reported on \cite{pgeg}.  The continuous-state description considers the rotation angle as a continuous random variable instead of supposing the angle to jump by a finite amount at each reactive event.  Accordingly, the continuous-state description allows us to incorporate mechanical aspects in the modeling which should otherwise be assumed in a discrete-state model, as explained here above.  Therefore, the continuous-state description provides the dependence of the rotational velocity on both the friction $\zeta$ and the external torque $\tau$ and is thus more complete in this regard.  Consequently, the dependence of the reaction constants of the discrete-state model on friction and external torque can be determined by comparison with the results of the simulation of our continuous-state model \cite{pgeg}, which is the purpose of the present section.

\subsection{Fokker-Planck equation description}
\label{Fokk}

In the continuous-state model \cite{pgeg}, the system is found at a given time $t$ in one of the six chemical states $\sigma=1,2,...,6$ and the $\gamma$-shaft at an angle $0\leq\theta<2\pi$. 
There are six chemical states because the three $\beta$-subunits can be either empty or occupied by a molecule of ATP or by the products ADP and P$_{\rm i}$ of hydrolysis.

Consequently, the system is described by six probability densities $p_{\sigma}(\theta,t)$ normalized according to $\sum_{\sigma=1}^6\int_0^{2\pi} p_{\sigma}(\theta,t)d\theta=1$.
The time evolution of the probability densities is ruled by a set of six Fokker-Planck equations coupled  together by the terms describing the random jumps between the chemical states $\sigma$ due to the two chemical reactions (ATP binding and release of the products ADP and P$_{i}$) with their corresponding reversed reactions \cite{pgeg}:
\be
\partial_t \; p_{\sigma}(\theta,t) + \partial_{\theta}\, J_{\sigma}(\theta,t)  = \sum_{\rho=1,2} \sum_{\sigma'(\neq\sigma)} 
\left[ p_{\sigma'}(\theta,t) \; w_{\rho,\sigma'\to\sigma}(\theta) - 
p_{\sigma}(\theta,t) \; w_{-\rho,\sigma\to\sigma'}(\theta) \right]
\label{FP.eqs}
\ee
where the probability current densities are given by
 \be
 J_{\sigma}(\theta,t) = -D \; \partial_{\theta}\, p_{\sigma}(\theta,t) + \frac{1}{\zeta}\left[-\partial_{\theta}U_{\sigma}(\theta)+\tau\right] p_{\sigma}(\theta,t)
 \label{fpflux}
 \ee

The diffusion coefficient $D$ can be expressed in terms of the friction coefficient $\zeta$ according to Einstein's relation $D=k_{\rm B}T/\zeta$.  The friction coefficient $\zeta$ can be evaluated for a bead attached to the $\gamma$-shaft, a bead duplex, or a cylindrical filament, as presented elsewhere \cite{N410,pgeg,KAI04,HGH94}. In the case of a bead of radius $r$ attached off-axis with its distance at a distance $x=r\sin\alpha$ from the rotation axis, the friction coefficient is given by
\be
\zeta =2 \pi \eta r^{3} \left(4+3 \sin^2\alpha\right) 
\label{coefffrict}
\ee
with the water viscosity $\eta = 10^{-9}$ pN s nm$^{-2}$ and $\alpha=\pi/6$ \cite{pgeg,KAI04}.

When the motor is in the chemical state $\sigma$, the $\gamma$-shaft is submitted to the external torque $\tau$ and the internal torque $-\partial_{\theta}U_{\sigma}$ due to the free-energy potential $U_{\sigma}(\theta)$ of the motor with its $\gamma$-shaft at the angle $\theta$.  Applying an external torque to the motor has the effect of tilting the potentials into $U_{\sigma}(\theta)-\tau \theta$ which eases the rotation or makes it harder, depending on the sign of $\tau$.

\subsection{Coarse graining into a discrete-state model}

The correspondence with the discrete-state description in terms of the master equation (\ref{meq}) can in principle be established by coarse graining the continuous angle into discrete states.
These discrete states correspond to the angular intervals 
$\theta_{\sigma}<\theta<\theta_{\sigma}+2\pi/3$ where the $\gamma$-shaft spends most 
of its time while in the chemical state $\sigma$.  Accordingly, the probabilities ruled by the master equation (\ref{meq}) are related to the probability densities of the continuous-state description (\ref{FP.eqs}) according to
\be
P_{\sigma}(t) = \int_{\theta_{\sigma}}^{\theta_{\sigma}+2\pi/3} p_{\sigma}(\theta,t) \; d\theta
\label{coarse}
\ee
where the angular integral is carried out over the aforementioned intervals.  In this way, a fully discrete description could be inferred from the continuous-state description.  In general, this reduction from one description to the other by the aforementioned coarse graining leads to non-Markovian equations.  In the case where there is a net separation of time scales between the dwell times  
and the transit times between the discrete states, the non-Markovian effects may be negligible and a description in terms of a Markovian equation such as the master equation (\ref{meq}) may be obtained.  This is the situation we here consider.

In this case, the transition rates $W_{\rho}(\sigma'\vert\sigma)$ of the master equation (\ref{meq}) can be deduced from the solution of the Fokker-Planck equations (\ref{FP.eqs}).  We emphasize that the transition rates $W_{\rho}(\sigma'\vert\sigma)$ of the master equation (\ref{meq}) do not take the same values as the rates $w_{\rho,\sigma'\to\sigma}(\theta)$ appearing in the Fokker-Planck equations (\ref{FP.eqs}). Indeed, the latter ones depend on the angle although the former ones do not.  A method of deduction of the former from the latter with Eq. (\ref{coarse}) would show that the transitions rates of the discrete-state description are given by intergrals of the continuous-angle rates in Eq. (\ref{FP.eqs}) combined with the probability densities of the quasi-stationary dwell states corresponding to the substeps. In spite of their conceptual and numerical differences, the transition rates appearing in both descriptions are in correspondence as far as the chemical reaction they describe are concerned, as shown in Table \ref{compar}. 

\begin{table}[htdp]
\caption{Comparison of the transition rates of the discrete model ruled by the master equation (\ref{meq}) with those of the continuous model ruled by the Fokker-Planck equations (\ref{FP.eqs}).   The transition rates are in correspondence by the chemical reaction they describe: ATP binding for $\rho=+1$; ATP unbinding for $\rho=-1$; the release of ADP and P$_{\rm i}$ for $\rho=+2$; the binding of ADP and P$_{\rm i}$ for $\rho=-2$.  We notice that the rates of the continuous model depend on the angle $\theta$ of the $\gamma$-shaft.  This dependence is of Arrhenius type in terms of the inverse temperature $\beta=(k_{\rm B}T)^{-1}$ and the free-energy potentials
$U(\theta)$ and $\tilde{U}(\theta)$ (for respectively the empty and occupied catalytic sites) and $U^{\ddagger}(\theta)$ and $\tilde{U}^{\ddagger}(\theta)$ (for respectively the transition states of ATP binding or unbinding and of the release or binding of ADP and P$_{\rm i}$).  See Ref. \cite{pgeg} for the forms of these potentials and the numerical values of the parameters of our continuous model.}
\begin{center}
\begin{tabular}{|c|c|}
\hline
$W_{\rho}(\sigma'\vert\sigma)$ & $w_{\rho,\sigma'\to\sigma}(\theta)$\\
\hline
$k_{+1}[{\rm ATP}]$ & $k_{0}[{\rm ATP}]\exp \left\lbrace -\beta \left[U^{\ddagger}(\theta)-U(\theta)-G^{\circ}_{\rm ATP} \right] \right\rbrace$\\
$k_{-1}$ & $k_{0}\exp \left\lbrace -\beta \left[U^{\ddagger}(\theta)-\tilde{U}(\theta) \right]\right\rbrace$\\
$k_{+2}$ & $\tilde{k}_{0}\exp \left\lbrace -\beta \left[\tilde{U}^{\ddagger}(\theta)-\tilde{U}(\theta+\frac{2\pi}{3}) \right]\right\rbrace$\\
$k_{-2}[{\rm ADP}][{\rm P_i}]$ \ & \ $\tilde{k}_{0} [{\rm ADP}][{\rm P_i}]\exp \left\lbrace -\beta \left[\tilde{U}^{\ddagger}(\theta)-U(\theta)-G^{\circ}_{\rm ADP}-G^{\circ}_{\rm P_{i}} \right]\right\rbrace$\\
\hline
\end{tabular}
\end{center}
\label{compar}
\end{table}

We notice that the rates $w_{\rho,\sigma'\to\sigma}(\theta)$ appearing in the Fokker-Planck equations (\ref{FP.eqs}) only concern the chemical reactions and, therefore, do not depend on the friction coefficient $\zeta$ and the external torque $\tau$ which only enter in the current densities (\ref{fpflux}) appearing in the left-hand side of the Fokker-Planck equations (\ref{FP.eqs}).  In contrast, the reaction constants $k_{\rho}$ shown in Table \ref{compar} necessarily depend on the friction coefficient $\zeta$ and the external torque $\tau$, which otherwise would not appear in the discrete model given by the master equation (\ref{meq}).  Accordingly, we now need to determine the dependences $k_{\rho}(\zeta,\tau)$ of the discrete-state reaction constants on both friction and external torque, which is the purpose of the next subsection.

\subsection{Dependence of the reaction constants on friction and external torque}

The original and key feature of our work is the inclusion in the discrete-model reaction constants $k_{\rho}$ ($\rho=\pm1, \pm2$) of the mathematical dependence on friction and external torque.
This dependence expresses the coupling between the chemical and mechanical properties of the motor.

We determine the dependences of the reaction constants $k_{\rho}$ on both friction and external troque by fitting them to the simulations of our continuous-angle model \cite{pgeg}.  Since this latter has been fitted to experimental data, the present fitting procedure is comparable to a fitting to the experimental data of Ref. \cite{N410}.  We use the method presented in subsection \ref{derivConst}.
In this way, we evaluate successively the reaction constants thanks to Eqs. (\ref{k+2})-(\ref{k-1}) for different values of the external torque $\tau$ and the friction $\zeta$.

In this regard, an essential observation is that the reaction constants become independent of the friction coefficient $\zeta$ at low friction and decrease as the inverse of the friction at high friction: $k_{\rho}\propto\zeta^{-1}$.  The motor is functioning in a reaction-limited regime at low friction and in a friction-limited regime at high friction \cite{pgeg}.  At high friction, the substeps are no longer visible since the rotation is slow down by the friction, in which case the Fokker-Planck equations (\ref{FP.eqs}) suggest indeed that the rate constants should scale as $\zeta^{-1}$.

The crossover between the low- and high-friction regimes can be well described by giving the following analytical form to the reaction constants:
\begin{equation}
k_{\rho}(\zeta,\tau) = \frac{1}{{\rm e}^{a_{\rho}(\tau)}+{\rm e}^{b_{\rho}(\tau)}\zeta}
\label{k(z,t)}
\end{equation}
with $\rho=\pm1,\pm2$.
The coefficients of the function in the denominator are taken as exponentials in order to guarantee the positivity of the reaction constants, the friction coefficient $\zeta$ being always non-negative.
The coefficients ${\rm e}^{a_{\rho}(\tau)}$ can be determined in the low-friction regime and the coeffcients ${\rm e}^{b_{\rho}(\tau)}$ in the high-friction regime.

The dependence on the external torque $\tau$ appears in the functions $a_{\rho}(\tau)$ and $b_{\rho}(\tau)$, which are taken as expansions in powers of $\tau$ limited to the second order:
\bea
a_{\rho}(\tau)&=&a_{\rho}^{(0)}+a_{\rho}^{(1)}\tau+a_{\rho}^{(2)}\tau^{2}+O(\tau^{3}) \label{atau}\\
b_{\rho}(\tau)&=&b_{\rho}^{(0)}+b_{\rho}^{(1)}\tau+b_{\rho}^{(2)}\tau^{2}+O(\tau^{3}) \label{btau}
\eea
with $\rho=1,2$.
The coefficients of these expansions are fitted in intervals of values of the external torque which are typically $\vert\tau\vert < 20$~pN~nm.
The values of the coefficients of Eqs. (\ref{atau})-(\ref{btau}) for the reaction constants $k_{+1}(\zeta,\tau)$, $k_{+2}(\zeta,\tau)$, and $k_{-2}(\zeta,\tau)$ are given in Table \ref{tabla00}.

\begin{table}
\caption{\label{tabla00} Values of the coefficients of the Taylor expansions (\ref{atau})-(\ref{btau}) of the functions $a_{\rho}(\tau)$ and $b_{\rho}(\tau)$ giving the reaction constants $k_{+1}(\zeta,\tau)$, $k_{+2}(\zeta,\tau)$, and $k_{-2}(\zeta,\tau)$, according to Eq.~(\ref{k(z,t)}).}
\begin{center}
\begin{tabular}[t]{|ccccc|}
\hline
coefficient & $k_{+1}(\zeta,\tau)$ & $k_{+2}(\zeta,\tau)$ & $k_{-2}(\zeta,\tau)$ & units\\
\hline
$a_{\rho}^{(0)}$ &  $-16.952$          & $-5.973$ & $-19.382$ & -\\
$a_{\rho}^{(1)}$ &  $9.8 \, 10^{-4}$ & $1.7 \, 10^{-4}$ & $1.29 \, 10^{-1}$ &(pN nm)$^{-1}$\\
$a_{\rho}^{(2)}$ &  $5.8 \, 10^{-4}$ & $1.0 \, 10^{-3}$ & $2.8 \, 10^{-4}$ &(pN nm)$^{-2}$\\
\hline
$b_{\rho}^{(0)}$ &  $-16.352$          & $-2.960$ & $-18.338$ &-\\
$b_{\rho}^{(1)}$ &  $-6.6 \, 10^{-2}$ & $-2.7 \, 10^{-2}$  & $5.9 \, 10^{-3}$ &(pN nm)$^{-1}$\\
$b_{\rho}^{(2)}$ &  $1.0 \, 10^{-3}$ &  $3.6 \, 10^{-4}$ & $-2.1 \, 10^{-4}$ & (pN nm)$^{-2}$\\
\hline
\end{tabular}
\end{center}
\end{table}

The last constant for ATP unbinding is finally obtained by using Eq. (\ref{k-1}) as
\begin{equation}
k_{-1}(\zeta,\tau) = \frac{k_{+1}(\zeta,\tau)\, k_{+2}(\zeta,\tau)}{k_{-2}(\zeta,\tau)} \exp \frac{1}{k_{\rm B}T}\left(\Delta G^{0}-\frac{2\pi}{3}\tau \right)
\label{k-1func}
\end{equation}
with $\Delta G^{0}=-50$~pN~nm.

%%%%%%%%%%%%%%%%%%%%%%%%%%%%%%%%%%%%%

\section{Properties of the F$_1$ motor}
\label{Properties}

The comparison between the discrete and continuous descriptions sheds a new light on the properties of the F$_1$ motor.  On the one hand, the continuous-angle model reproduces the experimental observations of Ref. \cite{N410} and its simulation can test the assumptions of the discrete-state model, in particular, the assumption of tight coupling.  On the other hand, the discrete-state model can be treated analytically thanks to its simplicity.  In this way, the fitting of the discrete-state model to the continuous one allows us to investigate more closely the properties of the F$_1$ motor in the regime of validity of the discrete-state model.  

\subsection{Tight versus loose chemomechanical coupling}

\begin{figure}[htbp]
\begin{center}
\includegraphics[width=8cm]{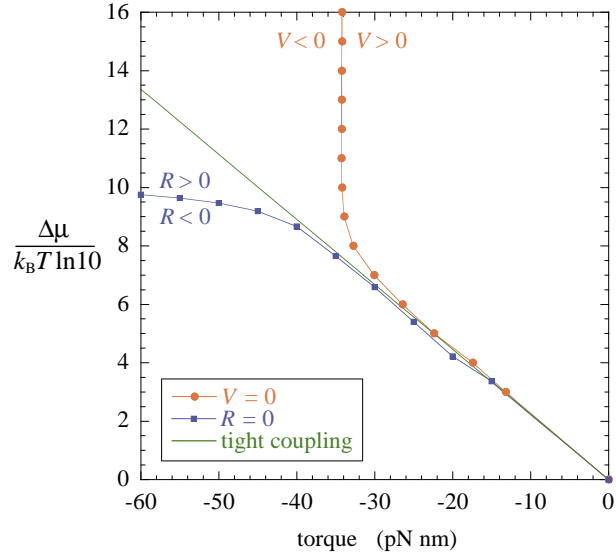}
\caption{Chemical potential difference $\Delta\mu$ in units of $k_{\rm B}T\ln 10$ versus the external torque $\tau$ for the situations where the rotation rate $V$ (circles) and the ATP consumption rate $R$ (squares) vanish in the continuous  model (\ref{FP.eqs}) and compared with the straight line $\Delta\mu=-2\pi\tau/3$ where the chemomechanical affinity (\ref{aff}) vanishes, $A=0$.  The concentrations are fixed according to $[{\rm ATP}]=4.9 \, 10^{0.8a-11}$~M and $[{\rm ADP}][{\rm P_i}]=10^{-0.2a-5}$~M, in terms of the quantity $a=\Delta\mu/(k_{\rm B}T\ln 10)$.
The bead attached to the $\gamma$-shaft has the diameter $d=2r=80$~nm and the temperature is of 23~degrees Celsius.  The torque where $V=0$ is called the stall torque.  The determination of the curves $V=0$ and $R=0$ is difficult close to the thermodynamic equilibrium point $(\tau=0,\Delta\mu=0)$ because both the rotation rate $V$ and the ATP consumption rate $R$ are very small in this region, which explains the absence of dots close to the origin.}
\label{fig1}
\end{center}
\end{figure}

In order to determine the regime of tight coupling between the chemistry and the mechanics of the F$_1$ motor, both the angular velocity $V$ and the ATP consumption rate $R$ have been simulated with the continuous-angle model (\ref{FP.eqs}) for different values of the external torque $\tau$ and chemical potential difference $\Delta\mu$, which are the corresponding affinities.  We depict in Fig. \ref{fig1} the curves in the plane $(\tau,\Delta\mu)$ where either the velocity vanishes, $V=0$, or the ATP consumption rate vanishes, $R=0$.  The value of the external torque where the velocity vanishes is the so-called stalling torque.  Since the variations of the chemical potential difference $\Delta\mu$ can only be one dimensional in such a representation, we take a combined variation of the concentrations of the difference species as explained in the caption of Fig. \ref{fig1}.  The two curves $V=0$ and $R=0$ intersect at the origin $(\tau=0,\Delta\mu=0)$ which is the thermodynamic equilibrium point.
We notice that the curve $V=0$ is above the curve $R=0$ in the plane of the chemical potential difference $\Delta\mu$ versus the torque as it should according to the second law of thermodynamics (\ref{diSdt}).

In the tight-coupling regime, the condition (\ref{V_tc}) should hold, which implies that the mechanical and chemical affinities $\tau$ and $\Delta\mu$ are no longer independent but should combined into the unique chemomechanical affinity (\ref{aff}).  In this regime, the vanishing of the rates, $V=0$ and $R=0$, should thus occur on the curve where the chemomechanical affinity (\ref{aff}) vanishes, i.e., along the straight line
\be
\Delta\mu = - \frac{2\pi}{3} \tau
\label{line}
\ee
This is observed in Fig. \ref{fig1} for values of the external torque extending from zero down to about $\tau \simeq -30$~pN~nm and for chemical potential difference from zero up to $\Delta\mu\simeq 14\, k_{\rm B}T$, which delimits the zone where the tight-coupling assumption is satisfied.  Outside this zone for higher values of $\Delta\mu$ or lower values of $\tau$, the F$_1$ motor is no longer functioning in the tight-coupling regime and the chemomechanical coupling becomes loose \cite{OH86}.

Since the coincidence of the curves $V=0$ and $R=0$ along the straight line (\ref{line}) is a feature of the discrete-state model (\ref{probaevo}), we may expect it describes the F$_1$ motor in the aforementioned zone of tight coupling.

\subsection{Rotation rate versus ATP concentration}

\begin{figure}[htbp]
\begin{center}
\includegraphics[width=7cm]{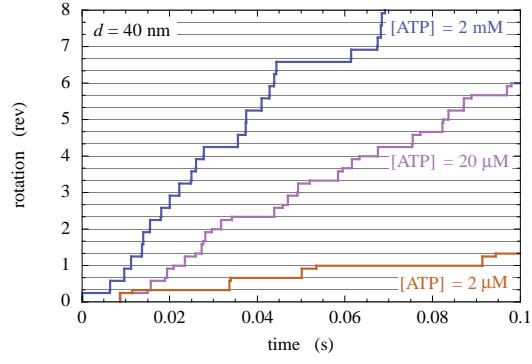}
\caption{Simulation of random trajectories of the discrete-state model for [ADP][P$_{i}$]=0, a temperature of 23 degrees Celsius, a bead of diameter $d=2r=40$~nm, and a zero external torque.}
\label{fig2}
\end{center}
\end{figure}

Random trajectories of the discrete model can be simulated thanks to Gillespie's numerical algorithm \cite{gillespie1,gillespie2}. Examples of random trajectories are depicted in Fig. \ref{fig2} for different values of ATP concentration, illustrating the Michaelis-Menten kinetics described by Eq. (\ref{v_MM2}). At low concentrations $[{\rm ATP}]\ll K_{\rm M}\simeq 17\, \mu{\rm M}$, the motor is essentially waiting for the arrival of new ATP molecules with its $\gamma$-shaft at some angle $120^{\circ}n$ ($n$ integer).
Instead, at high concentrations $[{\rm ATP}]\gg K_{\rm M}$, the rotation is limited by the release of the hydrolysis products ADP and P$_{\rm i}$ from its catalytic sites with its $\gamma$-shaft at some angle $90^{\circ}+120^{\circ}n$ ($n$ integer), as clearly seen in Fig. \ref{fig2}.  In this respect, the random trajectories of the discrete model reproduce the jumps by $90^{\circ}$ and $30^{\circ}$ corresponding to the substeps experimentally observed in Ref. \cite{N410}. We notice, however, that the discrete model cannot reproduce the small-amplitude fluctuations around the dwell angles as the continuous-angle model does  \cite{pgeg}.

\begin{figure}
\includegraphics[width=7cm]{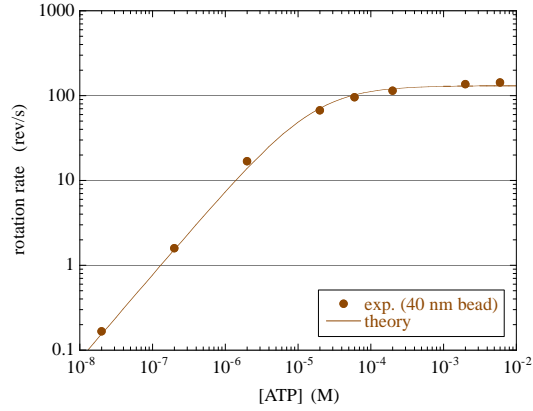}
\caption{Mean rotation rate of the $\gamma$-shaft of F$_{1}$ in revolutions per second, versus the ATP concentration [ATP] in mole per liter for [ADP][P$_{i}$]=0.
In accordance with the experimental setup \cite{N410}, the diameter of the bead is $d=2r=40$ nm, the temperature is of 23 degrees Celsius, and the external torque is zero.
The circles are the experimental data from Ref. \cite{N410}.
The solid line is the result of numerical simulation of the two-state model.}
\label{fig3}
\end{figure}

The crossover between the regime at low ATP concentration and the saturation regime of the Michaelis-Menten kinetics is seen in Fig. \ref{fig3} where we directly compare Eq. (\ref{v_MM2}) of the discrete model with experimental data from Ref. \cite{N410}.  At low ATP concentrations, the rotation rate is proportional to the ATP concentration while the rotation rate reaches its maximum value of about 130 rev/s in the saturation regime.

\begin{figure}
\includegraphics[width=7cm]{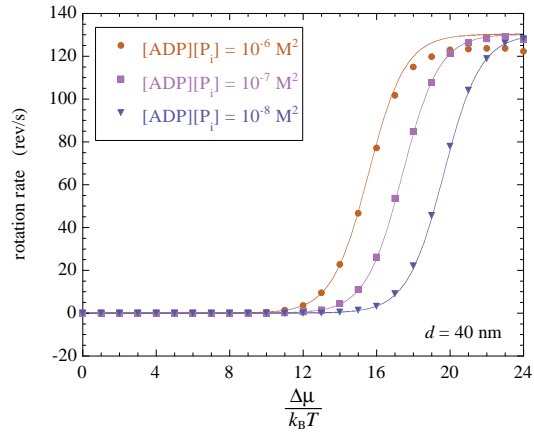}
\caption{Mean rotation rate versus the chemical potential difference $\Delta\mu$ in units of the thermal energy $k_{\rm B}T$. The thermodynamic equilibrium corresponds to $\Delta\mu=0$.  The ATP concentration is given in terms of the chemical potential difference by
$[{\rm ATP}]=[{\rm ADP}][{\rm P_{i}}]\exp[(\Delta\mu-\Delta\mu^{0})/(k_{\rm B}T)]\simeq 4.9\, 10^{-6}\, {\rm M}^{-1} [{\rm ADP}][{\rm P_{i}}]\exp[\Delta\mu/(k_{\rm B}T)]$ 
since $\Delta\mu^{0}=-\Delta G^{0}=50$~pN~nm.
The  results of the discrete model (solid lines) are compared with the continuous model (dots)
 for three different values of [ADP][P$_{i}$].
The diameter of the bead is $d=2r=40$~nm, the temperature 23~degrees Celsius, 
and the external torque zero.}
\label{fig4}
\end{figure}

In order to appreciate the nonequilibrium thermodynamics of the molecular motor, it is interesting to depict the rotation rate as a function of the affinity (\ref{aff}) instead of the ATP concentration.  Indeed, the former is a substitute of the latter, as shown by expressing the concentrations in terms of the chemical potentials by Eq. (\ref{defPotchim}) and using the definitions of the chemomechanical affinity (\ref{aff}) and of the equilibrium constant (\ref{K_eq}) to get
\be
[{\rm ATP}] = K_{\rm eq} [{\rm ADP}][{\rm P_i}] \, {\rm e}^A
\label{ATP-A}
\ee
In this way, we recover the equilibrium relation (\ref{ratioFb}) between the concentrations with Eq. (\ref{K_eq}) in the thermodynamic equilibrium state $A=0$ corresponding to given concentrations of ADP and P$_{\rm i}$. Substituting Eq. (\ref{ATP-A}) into Eq. (\ref{ATP+ADP.kinetics}), we obtain the following expression for the rotation rate:
\be
V = \frac{ V_{\rm max} \, \left( {\rm e}^A-1\right)}{{\rm e}^A-1 + \frac{3V_{\rm max}}{L}}
\label{V-Aff}
\ee
with the constant
\be
L \equiv \frac{3V_{\rm max}K_{\rm eq} [{\rm ADP}][{\rm P_i}]}{(K_{\rm eq}+K_{\rm P}) [{\rm ADP}][{\rm P_i}]+K_{\rm M}}
\ee
This coefficient controls the linear response of the molecular motor because
\be
V \simeq \left\{
\begin{array}{lcc}
\frac{1}{3}LA  & \mbox{for} & A\ll 1 \\
V_{\rm max} & \mbox{for} & A\gg 1
\end{array}
\right.
\ee
The analytic form (\ref{V-Aff}) shows that the rotation rate depends on the thermodynamic force $A$ in a highly nonlinear way, in contrast to what  is often supposed.  The nonlinear dependence is very important as observed in Fig. \ref{fig4}.  The linear regime extends around the thermodynamic equilibrium point at $\Delta\mu=0$ where the function $V(A)$ is essentially flat because the linear-response coefficient takes the very small value $L\simeq 10^{-5} \, {\rm s}^{-1}$.  Since the affinity is about $A\simeq 21.4$ under the physiological conditions $[{\rm ATP}]\simeq 10^{-3}$~M, $[{\rm ADP}]\simeq 10^{-4}$~M, and $[{\rm P_i}]\simeq 10^{-3}$~M \cite{KAI04}, the rotation rate would take the extremely low value $V\simeq LA/3\simeq 6.5$~rev/day if the motor was functioning in the linear regime.
Remarkably, the nonlinear dependence of Eq. (\ref{V-Aff}) on the affinity $A$ allows the rotation rate to reach the maximum value $V_{\rm max} \simeq 130$~rev/s under physiological conditions.

\subsection{Rotation rate versus friction}

\begin{figure}
\includegraphics[width=9cm]{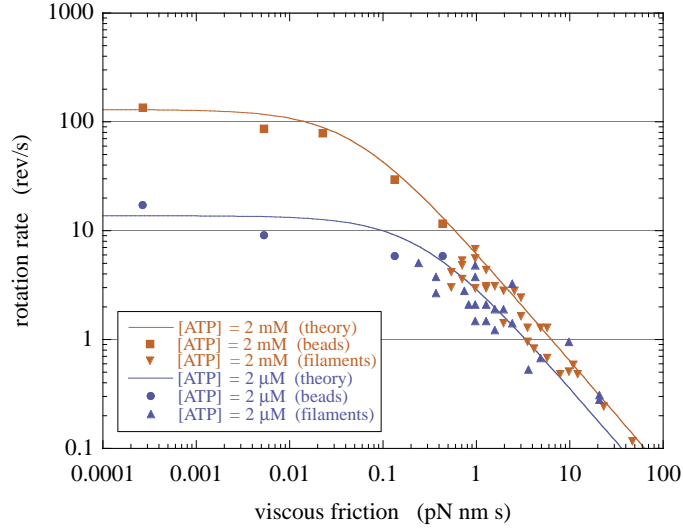}
\caption{The mean rotation rate in revolutions per second versus the viscous friction coeffcient $\zeta$ for a bead, a bead duplex, or a filament attached to the $\gamma$-shaft. The circles and triangles are the experimental data in Fig. 2 of Ref. \cite{N410} at [ATP] = 2 mM (squares and downward triangles) and [ATP] = 2 $\mu$M (circles and upward triangles). 
The squares and circles correspond to the single beads and bead duplexes, the triangles to the actin filaments.
The solid lines are the results of the present model with [ADP][P$_{\rm i}$] = 0, 
the temperature of 23 degrees Celsius, and zero external torque.}
\label{fig5}
\end{figure}

In Fig. \ref{fig5}, we show the effect of friction on the motor's velocity 
in the absence of ADP or P$_{\rm i}$. At low friction, the velocity saturates 
exhibing the reaction-limited regime. At high friction, we see the rapid decrease of the velocity with increasing friction in the friction-limited regime \cite{pgeg}.  
In this latter regime, the velocity decreases as the inverse of the friction coefficient $V\propto \zeta^{-1}$ in consistency with the analytical form (\ref{k(z,t)}) given to the reaction constants.  
A similar behavior is observed at [ATP] = 2 mM in the saturation regime of the Michaelis-Menten kinetics and at ATP concentrations lower than the Michaelis-Menten constant for [ATP] = 2 $\mu$M $< K_{\rm M} = 17 \, \mu$M. The agreement is as good as for the continuous-angle model \cite{pgeg}.

\subsection{Rotation in the presence of ATP hydrolysis products}

\begin{figure}
\includegraphics[width=8cm]{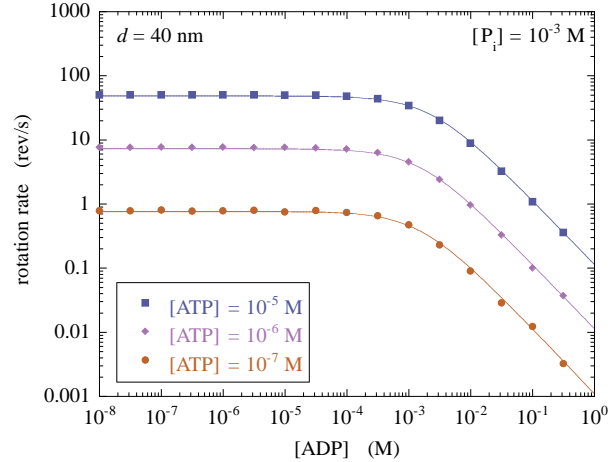}
\caption{Mean rotation rate of a bead of radius $r=20$~nm attached to the $\gamma$-shaft in revolutions per second versus [ADP] (in mole per liter) and varying ATP concentrations.
The concentration of P$_{\rm i}$ is $10^{-3}$ M and the temperature of 23~degrees Celsius, 
and the external torque vanishing.  The simulations of the continuous-angle model (squares, diamonds, and circles) are compared with the analytical expression (\ref{ATP+ADP.kinetics}) for the discrete model of Table \ref{tabla00} (lines).}
\label{fig6}
\end{figure}

In the presence of ADP and $P_{\rm i}$ in the environment of the motor, Eq. (\ref{ATP+ADP.kinetics}) shows that the rotation rate decreases, as expected since these products tend to counteract ATP hydrolysis that is powering the motor. This phenomenon is known as ADP inhibition \cite{N410,KAI04}.  There are two possible causes of the decrease of rotation rate if the concentrations of ADP and P$_{\rm i}$ are positive: (1) the term in the numerator of Eq. (\ref{ATP+ADP.kinetics}) where the equilibrium constant $K_{\rm eq}$ multiplies [ADP][P$_{\rm i}$]; (2) the term in the denominator of (\ref{ATP+ADP.kinetics}) where the constant $K_{\rm P}$ multiplies [ADP][P$_{\rm i}$].  As explained in subsection \ref{derivConst} and can be checked with Table \ref{tabla00}, the inequality (\ref{P>eq}) holds by several orders of magnitude so that the main cause of the decrease of the rotation rate is the term in the denominator of Eq. (\ref{ATP+ADP.kinetics}) due to
the reaction $\rho=-2$ of binding of ADP and P$_{\rm i}$ to the catalytic sites of the F$_1$ motor.

As observed in Fig. \ref{fig6} which compares the continuous and discrete models, this effect manifests itself above millimolar concentrations of ADP if inorganic phosphate is in millimolar concentration.  We notice that the decrease of the rotation rate goes as the inverse of the concentrations of ADP and P$_{\rm i}$, $V\propto ([{\rm ADP}][{\rm P_i}])^{-1}$ as described by Eq. (\ref{ATP+ADP.kinetics}) given the fact that the inequality (\ref{P>eq}) holds.

\subsection{Dependence on the external torque}

\begin{figure}
\includegraphics[width=8cm]{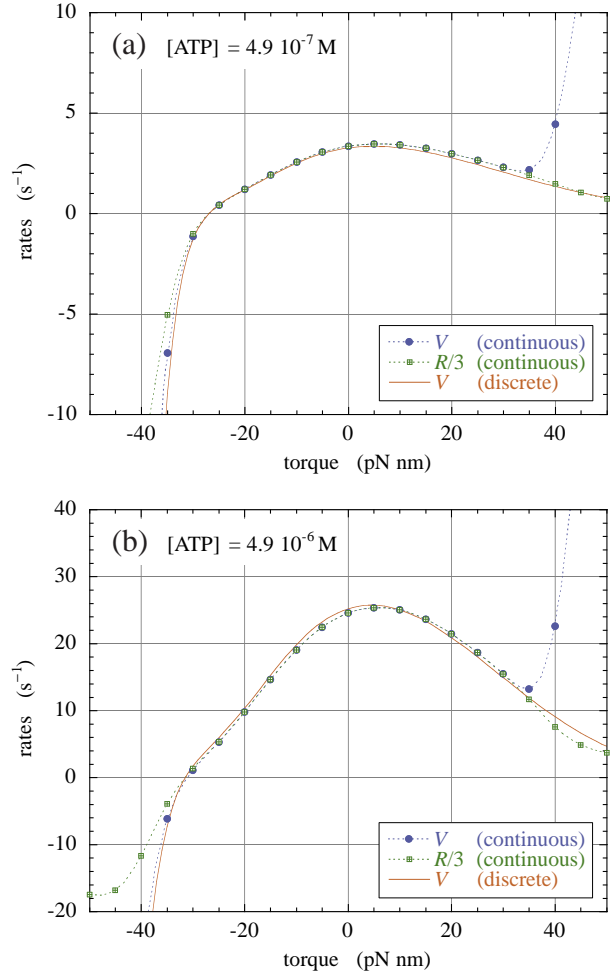}
\caption{Mean rotation rate $V$ of the $\gamma$-shaft of F$_{1}$ in revolutions per second versus the external torque for (a) [ATP]~=~$4.9 \, 10^{-7}$~M and (b) [ATP]~=~$4.9 \, 10^{-6}$~M (circles for the continuous-state model and solid line for the discrete-state model).
The other concentrations are fixed to [ADP]~=~$10^{-4}$~M and [P$_{\rm i}$]~=~$10^{-3}$~M.
The squares show the reaction rate $R$ divided by three in order to display the regime of tight coupling where $V=R/3$. 
The diameter of the bead is $d=2r=160$~nm and the temperature of 23~degrees Celsius.}
\label{fig7}
\end{figure}

Figure \ref{fig7} depicts the dependence of the rotation and ATP consumption rates on the external torque $\tau$ for both the continuous-angle and discrete-state models, showing their agreement in the range of validity of tight coupling.  This range of validity is observed in Fig. \ref{fig1} to extend down to $\tau\simeq -30$~pN~nm. The reason is that the mechanical motion becomes decoupled from the chemical reactions if the external torque is too much tilting the internal free-energy potential surfaces of the motor. For vanishing or moderate values of the external torque, the potential surfaces have barriers which stop mechanical motion and allow for jumps between potential surfaces corresponding to different chemical states of the molecular motor, coupling in this way 
the mechanical motion to the chemical reactions.
Beyond some threshold for the external torque, 
these barriers disappear causing the decoupling between mechanics and chemistry \cite{pgeg}.
This decoupling may happen at both negative and positive thresholds for the external torque, as seen in Fig. \ref{fig7} for the continuous-angle model.  The tight-coupling condition $V=R/3$ is only observed in the interval $\vert \tau\vert < 30$~pN~nm for the continuous-angle model, although it always holds by assumption for the discrete-state model.  Therefore, the range of validity of the discrete model is restricted to the interval $\vert \tau\vert < 30$~pN~nm of values of the external torque, as already discussed about Fig. \ref{fig1}.

 Figure \ref{fig7} also shows the remarkable feature of the discrete model to reproduce the phenomenon of stalling torque that the mean rotation rate can be stopped at a negative critical value of the external torque opposing the rotational motion.  Indeed, the value of the external torque where both the rotation and ATP consumption rates vanish can be obtained in the discrete model by Eq. (\ref{V-Aff}), which shows that the chemomechanical affinity should vanish at the stalling torque.  Whereupon, we recover the condition (\ref{line}) giving the stalling torque as
 \be
 \tau_{\rm stall} = -\frac{3}{2\pi} \Delta\mu =- \frac{3}{2\pi}\left( \Delta \mu^0 + k_{\rm B}T \ln \frac{[{\rm ATP}]}{[{\rm ADP}][{\rm P_i}]}\right)
 \label{tau_stall}
 \ee
 in the tight-coupling regime.  For $[{\rm ADP}][{\rm P_i}]=10^{-7}\, {\rm M}^2$, the chemical potential difference takes the values $\Delta\mu=56.5$~pN~nm and $\Delta\mu=65.9$~pN~nm, for respectively $[{\rm ATP}]=4.9 \, 10^{-7}$~M and $[{\rm ATP}]=4.9 \, 10^{-6}$~M. Hence, Eq. (\ref{tau_stall}) gives respectively the stalling torques $\tau_{\rm stall} =-27.0$~pN~nm and 
$\tau_{\rm stall} =-31.5$~pN~nm, as indeed observed in Figs. \ref{fig7}a and \ref{fig7}b.
 
\subsection{Chemical and mechanical efficiencies}

Under a negative external torque $\tau <0$,  the F$_1$ motor can synthesize ATP, 
in which case the ATP consumption rate as well as the rotation rate are negative, $R<0$ and $V<0$.
In this regime of ATP synthesis, the chemical efficiency is defined as the ratio of the free energy stored in the synthesized ATP over the mechanical power due to the external torque:
\be
\eta_{\rm c} \equiv - \frac{ \Delta\mu \, R}{2\pi \tau \, V}
\label{eta_c}
\ee
such that $0\leq \eta_{\rm c} \leq 1$ \cite{rmp}.

A mechanical efficiency can similarly be defined in the regime where the rotation is powered by ATP as the inverse of the chemical efficiency \cite{rmp}:
\be
\eta_{\rm m} \equiv - \frac{2\pi \tau \, V}{ \Delta\mu \, R}
\label{eta_m}
\ee
The mechanical efficiency satisfies $0\leq \eta_{\rm m} \leq 1$ in the regime where the external torque is still non-positive while both the rotation rate and the ATP consumption rates are positive, $V>0$ and $R>0$.  

\begin{figure}
\includegraphics[width=8cm]{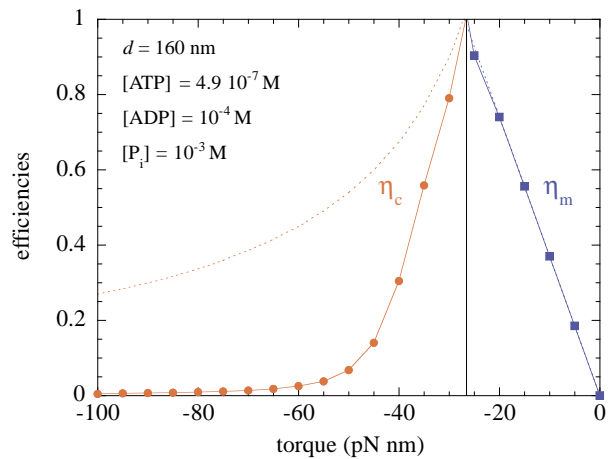}
\caption{Chemical efficiency (\ref{eta_c}) and mechanical efficiency (\ref{eta_m}) versus the external torque $\tau$ in the continuous-state model (respectively circles and squares joined by a solid line) and compared with the prediction (\ref{eta_c+eta_m_tc}) of tight coupling (dashed lines).  
The vertical solid line indicates the stalling torque at $\tau=-27.0$~pN~nm.
The concentrations are [ATP]~=~$4.9 \, 10^{-7}$~M, [ADP]~=~$10^{-4}$~M, and [P$_{\rm i}$]~=~$10^{-3}$~M. The diameter of the bead is $d=2r=160$~nm and the temperature of 23~degrees Celsius. These conditions are identical as in Fig. \ref{fig7}a.  The predictions of tight coupling (dashed lines) are respectively $\eta_{\rm c}=\tau_{\rm stall}/\tau$ for $\tau<\tau_{\rm stall}$, and $\eta_{\rm m}=\tau/\tau_{\rm stall}$ for $\tau_{\rm stall}<\tau<0$, with $\tau_{\rm stall}=-27.0$~pN~nm.}
\label{fig8}
\end{figure}

It is known \cite{rmp} that the chemical and mechanical efficiencies reach their maximum values under different conditions as shown in the Appendix. In the tight-coupling regime where $R=3V$, these conditions coincide and the chemical and mechanical efficiencies (\ref{eta_c}) and (\ref{eta_m}) become
\be
\mbox{tight coupling:} \qquad \eta_{\rm c} = \frac{1}{\eta_{\rm m}} = - \frac{3\Delta\mu}{2\pi\tau}
\label{eta_c+eta_m_tc}
\ee
in agreement with Eq. (\ref{eff_tc}) derived in the Appendix from the assumption of linear response.
In the tight-coupling regime, the chemical and mechanical efficiencies can reach the maximal unit value at the stalling torque where Eq. (\ref{tau_stall}) holds.  This remarkable result is observed in Fig. \ref{fig8} depicting the chemical and mechanical efficiencies versus the external torque under conditions corresponding to the chemical potential difference $\Delta\mu=56.5$~pN~nm.  

In the interval $\tau_{\rm stall}=-27.0$~pN~nm $< \tau <0$, the external torque is opposed to the mean rotation rate but the motor consumes ATP, so that the mechanical efficiency (\ref{eta_m}) is postive.  In this interval, the coupling is tight so that the mechanical efficiency computed with the continuous-angle model (squares) agrees with the prediction (\ref{eta_c+eta_m_tc}) of tight coupling which holds for the discrete model.  Indeed, the mechanical efficiency reaches the unit value at the stalling torque marked by the vertical line.  

On the other hand, the motor synthesizes ATP under the action of the external torque for $\tau <\tau_{\rm stall}=-27.0$~pN~nm, where  the velocity becomes negative and the chemical efficiency (\ref{eta_c}) positive.  As explained above, the motor is no longer in a regime of tight coupling below the stalling torque so that the chemical efficiency computed with the continuous-angle model differs from the prediction (\ref{eta_c+eta_m_tc}) of tight coupling. It is only at the stalling torque that both agree while they reach the unit value.  However, as seen in Fig. \ref{fig8}, the actual chemical efficiency is lower than possible by tight coupling because of the loose chemomechanical coupling below the stalling torque.

%%%%%%%%%%%%%%%%%%%%%%%%%%%%%%%%%%%%%

\section{Conclusions}
\label{Conclusions}

In the present paper, we have studied one of the simplest possible stochastic processes describing the stochastic chemomechanics of the F$_1$-ATPase molecular motor, as observed in Ref. \cite{N410}.  This description considers the two discrete states corresponding to the steps and substeps observed in the rotary motion of the $\gamma$-shaft of F$_{1}$ \cite{N410}. This two-state discrete model is based on the master equation ruling the time evolution of the probabilities of the two discrete states.  The model is analytically solvable, which provides us with an interesting insight in our understanding of the behavior of the F$_1$ motor.  The discrete model is set up by fitting its rate coefficients to simulations with our previously continuous-angle model \cite{pgeg}.  Since this latter was itself fitted to the observations of Ref. \cite{N410}, both the discrete and continuous models describe the same experimental system.

The comparison between the discrete-state and continuous-angle descriptions reveals important properties of the F$_1$ motor. Indeed, the discrete-state model presupposes that the mechanical motion and the chemical reactions powering the motor are tightly coupled, although the continuous-angle model does not.  Accordingly, the comparison between both models clearly reveal the regime of tight coupling.  Thanks to the analytic tractability of the discrete-state model, the consequences of the transition between the tight- and loose- coupling regimes can be understood for the kinetic and thermodynamic properties of the F$_1$ motor.

An important difference between the continuous and discrete models is that mechanical properties such as the friction of the object attached to the $\gamma$-shaft in the liquid surrounding the motor or the external torque acting on the shaft explicitly appear in the Fokker-Planck equations defining the continuous model, but does not appear in the master equation of the discrete model.  Indeed, this latter is defined by its transition rates and its rate constants although the former also contains a biased diffusive part involving both the friction coefficient $\zeta$ and the external torque $\tau$.
In order to describe these mechanical properties in the discrete-state model, we thus have to give a dependence on the friction coefficient $\zeta$ and the external torque $\tau$ in the rate constants, which are also the the constants of the reactions of the F$_1$ motor with ATP, ADP, and P$_{\rm i}$.
Accordingly, the discrete-state model is essentially a model of the chemical kinetics of the F$_1$ motor with reaction constants $k_{\rho}(\zeta,\tau)$ depending on mechanical parameters.  This dependence is fitted to numerical simulations of the continuous-angle model.  This fitting is performed in the tight-coupling regime where the discrete-state model is supposed to correspond to the continuous-angle model.  In this way, the two-state model is completed by including the dependences on both the chemical and mechanical control parameters.  The random time evolution of the discrete two-state model can be simulated thanks to Gillespie's numerical algorithm \cite{gillespie1,gillespie2}.

In the discrete-state model, the tight coupling between mechanics and chemistry implies that the motor is driven out of equilibrium by the unique chemomechanical affinity (\ref{aff}) combing the external torque with the chemical potential difference of ATP hydrolysis.  Thermodynamic considerations shows that the chemical equilibrium is displaced by the external torque acting on the motor if this latter strictly functions in the tight-coupling regime.  

Thanks to the solvability of the two-state model, the stationary solutions of the master equation can be exactly deduced, allowing us to show analytically that the mean rotation rate obeys a Michaelis-Menten kinetics with respect to the ATP concentration.  Moreover, the analytical formula (\ref{ATP+ADP.kinetics}) is obtained for the rotation rate in the presence of both ATP and the products of ATP hydrolysis, i.e., ADP and P$_{\rm i}$.

The highly nonlinear dependence of the mean rotation rate of the $\gamma$-shaft (\ref{V-Aff}) on the chemomechanical affinity (\ref{aff}) shows that the F$_1$ motor is not functioning in the linear-response regime defined by Onsager's linear-response coefficents, but instead typically runs in a nonlinear-response regime which is more the feature of far-from-equilibrium systems 
than of close-to-equilibrium systems. This remarkable property allows the motor to rotate under physiological conditions at about 130 rev/s although it would be a million times slower if the motor was functioning in a linear-response regime.

Furthermore, the crossover between the reaction-limited and friction-limited regimes is well described by the two-state model as the friction coefficient $\zeta$ is increased.  Although, the mean rotation rate is nealry independent of the friction coefficient in the reaction-limited regime at low friction, it decreases as the inverse of the friction coefficient at high friction because the reaction rates (\ref{k(z,t)}) fitted to the continuous-angle model appropriately take this feature into account. 

We have also investigated the behavior of the F$_1$ motor in a surrounding filled with ADP and inorganic phosphate.  As shown by Eq. (\ref{ATP+ADP.kinetics}), the mean rotation rate decreases as the concentrations of ADP and P$_{\rm i}$ exceed a crossover value.  The reason is that the release of the products of ATP hydrolysis in the motor is counteracted by the reverse reaction of binding of the products to the catalytic sites, which has the effect of slowing down the motor.

The dependence of the rotation and ATP consumption rates on the external torque is also of great interest because it reveals the interval of values of the external torque where the tight-coupling condition holds.  It is in this regime that the discrete-state model provides a good description of the motor.  
This comparison shows that the stalling torque where the rotation and ATP consumption rates vanish is given in terms of the chemical potential difference of ATP hydrolysis and thus depend on the concentrations of the involved species according to Eq. (\ref{tau_stall}).

Moreover, the nonequilibrium thermodynamics of the F$_1$ motor has been developed, allowing us to study the chemical and mechanical efficiencies defined in Ref. \cite{rmp}. In the tight-coupling regime, these efficiencies can reach their maximal unit value near the stalling torque.  The coupling between mechanics and chemistry is so tight that the expected deviations between the stalling torque and the values of the torque where the efficiencies reach their maximum value turn out to be very small.
The coincidence of these different values of the external torque is the consequence of the high degree of  tight coupling achieved in the F$_1$ motor, but would not necessarily hold for a motor functioning in a loose-coupling regime.

In conclusion, the two-state model and its comparison with the continuous-angle model of our previous paper \cite{pgeg} is providing a powerful method to study the kinetic and thermodynamic properties of the F$_1$ motor and, especially, the coupling between its mechanics and chemistry.

\vspace{0.3cm} 

{\bf Acknowledgments.} 
This research is financially supported by the ``Communaut\'e fran\c caise de Belgique''
(contract ``Actions de Recherche Concert\'ees'' No.~04/09-312),
by the ``Fonds pour la Formation \`a la
Recherche dans l'Industrie et l'Agriculture" (F.~R.~I.~A. Belgium), and
the National Fund for Scientific Research (F.~N.~R.~S. Belgium, contract F.~R.~F.~C. No.~2.4577.04).

%%%%%%%%%%%%%%%%%%%%%%%%%%%%%%%%%%%%%%%%%%%%%%%%%% 

\appendix

\section{Thermodynamic relations between affinities and currents}

In this appendix, the nonequilibrium thermodynamics of the F$_1$ motor is presented in the linear regime very close to the equilibrium.

\subsection{The general case}

The molecular motor can be driven out of equilibrium by the two independent affinities which are the mechanical and the chemical affinities proportional respectively to the torque $\tau$ and the chemical potential difference $\Delta\mu$.  These affinities are the corresponding fluxes or currents can be defined as 
\bea
A_{\rm m} \equiv \frac{2\pi}{3}\frac{\tau}{k_{\rm B}T} \quad &\leftrightarrow& \quad J_{\rm m} \equiv 3V \\
A_{\rm c} \equiv \frac{\Delta\mu}{k_{\rm B}T} \quad &\leftrightarrow& \quad J_{\rm c} \equiv R 
\eea
in which case the thermodynamic entropy production (\ref{diSdt}) takes the following form:
\be
\frac{1}{k_{\rm B}} \frac{d_{\rm i}S}{dt} = A_{\rm m} J_{\rm m} + A_{\rm c} J_{\rm c} \geq 0
\label{diSdt_AJ}
\ee

In general, the currents are nonlinear functions of the affinities
\bea
J_{\rm m} &=& J_{\rm m}(A_{\rm m},A_{\rm c})\\
J_{\rm c} &=& J_{\rm c}(A_{\rm m},A_{\rm c})
\eea
which vanish at the thermodynamic equilibrium where the affinities vanish, $A_{\rm m}=A_{\rm c}=0$.
Close to equilibrium, the currents can be expanded in powers of the affinities, which defines the linear response coefficients $L_{ij}$ as
\bea
J_{\rm m} &=& L_{11}A_{\rm m}+L_{12}A_{\rm c} + O(2) \label{Jm-lin}\\
J_{\rm c} &=& L_{21}A_{\rm m}+L_{22}A_{\rm c} + O(2) \label{Jc-lin}
\eea
The microreversibility implies the Onsager reciprocity relation:
\be
L_{12}=L_{21}
\ee
and the non-negativity of the entropy production (\ref{diSdt_AJ}) the inequality
\be
L_{11}L_{22} \geq L_{12}^2 \qquad \mbox{with} \quad L_{11}, L_{22} \geq 0 
\label{ineq_2ndlaw}
\ee
The coupling between the mechanics and the chemistry is here possible because the motor is attached
to a solid support and keeps a fixed orientation.  The chemistry thus becomes vectorial as well as mechanical. Therefore, there is no contradiction with the Curie symmetry principle according to which scalar and vectorial processes cannot be coupled together \cite{P67,KP98}.  In isotropic media such as liquids, chemistry is scalar and cannot be coupled to mechanics which is vectorial.  This coupling becomes possible if the medium is anisotropic as in liquid crystals or in proteins attached to the surface of a solid.

In the plane of the affinities $(A_{\rm m},A_{\rm c})$, the curve where the velocity vanishes behaves around the origin as
\be
J_{\rm m}=3V=0: \qquad A_{\rm c} = - \frac{L_{11}}{L_{12}} A_{\rm m} + O(A_{\rm m}^2)
\label{V=0}
\ee
while the curves where the ATP consumption rate vanishes is given by
\be
J_{\rm c}=R=0: \qquad A_{\rm c} = - \frac{L_{21}}{L_{22}} A_{\rm m} + O(A_{\rm m}^2)
\label{R=0}
\ee
Near the origin, the curve (\ref{V=0}) has therefore a more negative slope than the curve (\ref{R=0})
as the consequence of the inequality (\ref{ineq_2ndlaw}) resulting from the second law of thermodynamics \cite{rmp}.

\subsection{The case of tight coupling}

In the case of a tight coupling between the mechanics and the chemistry of the molecular motor, these two curves have the same slope around the thermodynamic equilibrium point.  Indeed, the tight-coupling condition (\ref{V_tc}) reads
\be
\mbox{tight coupling:} \qquad J_{\rm m} = J_{\rm c} \equiv J
\label{unique_current}
\ee
so that the entropy production becomes
\be
\mbox{tight coupling:} \qquad \frac{1}{k_{\rm B}} \frac{d_{\rm i}S}{dt} = A J \geq 0
\label{diSdt_tc_AJ}
\ee
with the chemomechanical affinity (\ref{aff}), which here reads
\be
A \equiv A_{\rm m} + A_{\rm c}
\label{aff_AJ}
\ee
Therefore, the mechanical and chemical affinities are no longer independent in the tight-coupling regime where the chemomechanical affinity (\ref{aff_AJ}) becomes the unique nonequilibrium driving force.  Accordingly, the unique current (\ref{unique_current}) is a function of this unique affinity:
\be
J=J(A)=LA + O(A^2)
\ee
and all the linear response coefficients become related to each other by
\be
L \equiv L_{11}=L_{12}=L_{21}=L_{22}
\label{Ls_tc}
\ee
In this case, the inequality (\ref{ineq_2ndlaw}) reaches the equality:
\be
L_{11}L_{22} = L_{12}^2
\label{equal_tc}
\ee
which is also characteristic of tight coupling.

\subsection{The efficiencies}

The chemical and mechanical efficiencies (\ref{eta_c})-(\ref{eta_m}) can be expressed as follows in terms of the affinities and the currents:
\be
\eta_{\rm c} = - \frac{A_{\rm c}J_{\rm c}}{A_{\rm m}J_{\rm m}} = \frac{1}{\eta_{\rm m}}
\label{eff}
\ee

In the regime of linear response where the currents are the linear functions (\ref{Jm-lin})-(\ref{Jc-lin})  of the affinities, the efficiencies can be written in the form
\be
\eta_{\rm c} = - \frac{1-\varepsilon}{\alpha} \, \frac{\alpha +1}{\alpha+1-\varepsilon} = \frac{1}{\eta_{\rm m}}
\label{fn-eff}
\ee
in terms of the coefficient
\be
\alpha \equiv \frac{L_{12}}{L_{22}} \frac{A_{\rm m}}{A_{\rm c}}
\label{alpha}
\ee
and the constant
\be
\varepsilon \equiv 1-\frac{L_{12}^2}{L_{11}L_{22}}
\label{eps}
\ee
such that $0\leq \varepsilon \leq 1$ by the inequality (\ref{ineq_2ndlaw}).
Equations (\ref{fn-eff}) and (\ref{alpha}) show that the efficiencies only depend on the ratio of the affinities \cite{rmp}.  In their respective domains of variation, the chemical and mechanical efficiencies can reach the maximum value
\be
0 \leq \eta_{\rm max} \equiv \frac{1-\sqrt{\varepsilon}}{1+\sqrt{\varepsilon}} \leq 1
\label{eta_max}
\ee
This happens along different curves in the plane $(A_{\rm m},A_{\rm c})$ of the affinities with respect to the curves where the rotation and ATP consumption rates vanish since \cite{rmp}
\bea
\eta_{\rm m} = \eta_{\rm max} \quad &\mbox{if}& \quad \alpha = - 1 + \sqrt{\varepsilon} \label{-1+se}\\
J_{\rm m} = 3V=0 \quad &\mbox{if}& \quad \alpha = - 1 + \varepsilon \label{-1+e}\\
J_{\rm c} = R=0 \quad &\mbox{if}& \quad \alpha = - 1  \label{-1}\\
\eta_{\rm c} = \eta_{\rm max} \quad &\mbox{if}& \quad \alpha = - 1 - \sqrt{\varepsilon}  \label{-1-se}
\eea

In the tight-coupling limit, the coefficient (\ref{eps}) vanishes because of Eq. (\ref{equal_tc}), $\varepsilon=0$, so that the chemical and mechanical efficiencies take the value
\be
\eta_{\rm c} = \frac{1}{\eta_{\rm m}} =-\frac{1}{\alpha} = - \frac{A_{\rm c}}{A_{\rm m}} 
\label{eff_tc}
\ee
according to Eqs. (\ref{fn-eff}) and (\ref{Ls_tc}).  In this case, the four conditions (\ref{-1+se})-(\ref{-1-se}) coincide in $\alpha=-1$ where the chemical and mechanical efficiencies reach their maximal value which is equal to unity, $\eta_{\rm max}=1$. Therefore, tight coupling favors the optimization of the efficiencies.

%%%%%%%%%%%%%%%%%%%%%%%%%%%%%%%%%%%%%%%%%%%%%%%%%%%%%%%%%%%%%%%%%%%%%%%%%%%%%%

\end{document}